\pdfoutput=1

\documentclass[11pt]{article}

\usepackage{acl}

\usepackage{times}
\usepackage{latexsym}

\usepackage[T1]{fontenc}

\usepackage[utf8]{inputenc}

\usepackage{microtype}

\usepackage{inconsolata}

\usepackage{graphicx}
\usepackage{inconsolata}
\usepackage{graphicx}
\usepackage{amsmath}
\usepackage{adjustbox}
\usepackage{multirow}
\usepackage{svg}
\usepackage[normalem]{ulem}
\useunder{\uline}{\ul}{}
\usepackage{makecell}
\usepackage{tablefootnote}

\usepackage{xcolor}

\definecolor{colorhigh}{HTML}{9900FF}
\definecolor{colorlow}{HTML}{660000}

\title{Efficient and Versatile Model for Multilingual Information Retrieval of Islamic Text: Development and Deployment in Real-World Scenarios}

\author{Vera Pavlova$^{1,2}$, Mohammed Makhlouf$^{1}$ \\
  $^{1}$rttl labs \quad $^{2}$burevestnik.ai \\
  $^{1}$\texttt{v@rttl.ai, mm@rttl.ai} \quad $^{2}$\texttt{v@burevestnik.ai}
}

\begin{document}
\maketitle
\begin{abstract}
Despite recent advancements in Multilingual Information Retrieval (MLIR), a significant gap remains between research and practical deployment. Many studies assess MLIR performance in isolated settings, limiting their applicability to real-world scenarios.
In this work, we leverage the unique characteristics of the Quranic multilingual corpus to examine the optimal strategies to develop an ad-hoc IR system for the Islamic domain that is designed to satisfy users' information needs in multiple languages. 
We prepared eleven retrieval models employing four training approaches: monolingual, cross-lingual, translate-train-all, and a novel mixed method combining cross-lingual and monolingual techniques. 
Evaluation on an in-domain dataset demonstrates that the mixed approach achieves promising results across diverse retrieval scenarios. Furthermore, we provide a detailed analysis of how different training configurations affect the embedding space and their implications for multilingual retrieval effectiveness.
Finally, we discuss deployment considerations, emphasizing the cost-efficiency of deploying a single versatile, lightweight model for real-world MLIR applications. The system is deployed online\footnote{\url{https://rttl.ai/}}.

\end{abstract}

\section{Introduction}
MLIR is a challenging area of research that has seen significant advancements recently, mainly due to the use of large language models (LLMs) \citep{Nair2022TransferLA, Lawrie2022NeuralAT}. However, there remains a considerable gap between research efforts and the actual deployment of MLIR systems in real-world scenarios. Many studies show impressive results in controlled environments or benchmark datasets, but typically focus on evaluating the IR model in a specific setting. However, many real-world applications often require a combination of various search scenarios within a single IR system —be it multilingual, cross-lingual, or monolingual. One example of such an application is a retrieval task for the Holy Quran. 
Retrieving passages from the Holy Quran is uniquely challenging. With translations in over 100 languages, it offers a rich parallel collection of high-quality human translations \citep{bashir2023arabic}. This unique feature provides an excellent opportunity to explore the multilingual potential of retrieval models and eliminates the bottleneck of applying machine translation (MT), simplifying and streamlining the evaluation process.

This study examines training approaches for deploying a single retrieval model across diverse MLIR settings, enabling modern search capabilities in the Islamic domain. The goal is to help users efficiently locate relevant Quranic passages in multiple languages and access the cultural and religious heritage preserved within Islamic texts, serving both scholars and the general public.
We utilize the XLM-R\textsubscript{Base} model \citep{conneau-etal-2020-unsupervised}, a multilingual model trained for a general domain, as a backbone for retrieval. It is known that the performance of retrieval models typically deteriorates due to domain shift \citep{thakur-2021-BEIR, pavlova-2023-leveraging}. As a preliminary step, we conduct a brief domain adaptation of the XLM-R\textsubscript{Base} model using a small multilingual domain-specific corpus (approximately 100M words). This short round of pre-training resulted in significant performance improvements in retrieval tasks. Moreover, to reduce the model's size, we conduct language reduction on the XLM-R\textsubscript{Base} model, which allows us to eliminate languages that are not needed for the current deployment, resulting in more than a 50\% reduction in the model's size.
We prepare eleven retrieval models using this lightweight domain-specific multilingual large language model (MLLM) by applying four different training approaches: monolingual, cross-lingual, translate-train-all, and a proposed mixed approach that combines cross-lingual and monolingual techniques. Evaluation across monolingual, cross-lingual, and multilingual retrieval scenarios demonstrates that our proposed mixed training approach produces promising results in all settings. We conduct an in-depth analysis of the potential effects of different training configurations on the embedding space and their impact on multilingual retrieval. 
Additionally, we discuss the advantages of deploying a single lightweight model for various potential deployment scenarios.

Our main contributions are:
(1) We prepared eleven retrieval models trained using different approaches and conducted rigorous testing. This allowed us to evaluate how these models perform in various retrieval settings without limiting the evaluation to each model's specific training approach. 
(2) We propose a mixed training approach that achieves competitive performance across monolingual, cross-lingual, and multilingual retrieval scenarios.
(3) We deploy our model as a part of a free online multilingual
search tool designed to explore Quranic text in multiple languages.

\section{Preliminaries}
\label{sec:Preliminaries}
In this work, we use the term MLIR in its broadest sense. This includes monolingual IR in any language other than English, cross-lingual IR as a special case of MLIR, and MLIR itself, which enables the processing of queries in any language while retrieving relevant documents in multiple languages \citep{Oard-1996, Oard1998EvaluatingCT}.
We experimented with four languages: English, Arabic, Urdu, and Russian.
The choice of languages is motivated by the availability of the evaluation dataset and diversity.
Before we proceed to the details of the training approach and experiment, we briefly discuss the preparation of a lightweight domain-specific MLLM that serves as a backbone of the retrieval models.

MLLMs provide cross-lingual functionality but are heavy to deploy in low-resource settings due to their large size \citep{devlin-etal-2019-bert, Lample-XLM, conneau-etal-2020-unsupervised}.
Language reduction is a promising approach in the deployment environment \citep{abdaoui-etal-2020-load}. It decreases the model size by pruning only the embedding matrix and removing the languages that are not needed in deployment while preserving all encoder weights. We use XLM-R\textsubscript{Base} to perform language reduction and trim its size from 1.1 GB down to 481MB by keeping the vocabulary only of languages of interest: English, Arabic, Urdu, and Russian (XLM-R-4 model). 
We follow the technique of \citet{pavlova-makhlouf-2024-building}. The detailed steps and model comparison are listed in the Appendix ~\ref{sec:appendix}.

\subsection{Domain Adaptation of MLLM}
\label{sec: Domain Adaptation}

While it is relatively easy to find a small amount of data for unsupervised pre-training, the absence of domain-specific labeled data for downstream tasks is a common problem. By leveraging continued pre-training and the integration of new domain-specific vocabulary \citep{Lee-2019, Huang-ClinicalBERT, Gu-PubMedBERT, beltagy-etal-2019-scibert, pavlova-makhlouf-2023-bioptimus, pavlova-2025-multi}, we perform domain adaptation using a small corpus of 100 million words only. We combine a random subset of 50 million words from The Open Islamicate Texts Initiative (domain-specific corpus in Arabic) \citep{romanov2019openiti} with texts in English, Russian, and Urdu, mainly consisting of Tafseer and Hadith, also totaling 50 million words. We train a new Islamic tokenizer based on this corpus, add new domain-specific tokens to the existing vocabulary of the XLM-R-4 model, and continue a short round of pre-training on this assembled multilingual Islamic corpus (XLM-R-4-ID model). For more details on the hyperparameters, refer to Appendix~\ref{sec:appendix}. As we will show below, this short pre-training round significantly boosted model performance on retrieval tasks.

\begin{table}[t]
\begin{adjustbox}{width=\columnwidth,center}
\begin{tabular}{lccccc}
\hline
\textbf{Models}                       & \textbf{EN}    & \textbf{AR}    & \textbf{UR}    & \textbf{RU}    & \textbf{avg.} 
\\ \hline
\textbf{XLM-R-EN}             & 0.365          & 0.057          & 0.291          & 0.305          & 0.254
\\ \hline
\textbf{EN-monolingual}             & 0.377          & 0.430          & 0.373          & 0.339          & 0.380
\\
\textbf{AR-monolingual}             & {\ul 0.436}    & 0.416          & 0.400          & 0.337          & {\ul 0.397}     \\ \hline
\textbf{ENq-ARc}             & \textbf{0.441} & 0.358          & 0.381          & 0.333          & 0.378          \\
\textbf{ARq-ENc}             & 0.133          & 0.406          & {\ul 0.427}    & \textbf{0.401} & 0.342          \\ \hline
\textbf{ENq-Bic}             & 0.418          & {\ul 0.434}          & 0.368          & 0.314          & 0.384
 \\ 
\textbf{ARq-Bic}             & 0.358          & 0.377          & 0.405          & 0.332          & 0.368
\\
\textbf{Biq-ENc}             & 0.430          & \textbf{0.452}    & \textbf{0.454} & {\ul 0.365}  & \textbf{0.426} 
\\
\textbf{Biq-ARc}             & 0.417          & 0.389          & 0.422          & 0.361          & 0.397
\\
\textbf{Biq-Bic}             & 0.349          & 0.365          & 0.407          & 0.333          & 0.363
\\ \hline
\textbf{Bilingual-train-all} & 0.407          & 0.386          & 0.360          & 0.327          & 0.370          \\
\textbf{4lingual-train-all}  & 0.421          & 0.366          & 0.273          & 0.341          & 0.350
\\ \hline
\end{tabular}
\end{adjustbox}
\caption{Monolingual evaluation MRR@10.}
\label{tab:table 1}
\end{table}

\begin{table}[t]
\begin{adjustbox}{width=\columnwidth,center}
\begin{tabular}{lccccc}
\hline
\textbf{Models}                       & \textbf{EN}    & \textbf{AR}    & \textbf{UR}    & \textbf{RU}    & \textbf{avg.}  \\ \hline

\textbf{XLM-R-EN}             & 0.237          & 0.197          & 0.165          & 0.255          & 0.222          \\ \hline
\textbf{EN-monolingual}             & 0.293          & 0.345          & {\ul 0.324}    & 0.289          & 0.326          \\
\textbf{AR-monolingual}             & 0.329          & 0.336          & 0.313          & 0.265          & 0.328          \\ \hline
\textbf{ENq-ARc}             & 0.350          & 0.272          & 0.286          & 0.264          & 0.310          \\
\textbf{ARq-ENc}             & 0.112          & {\ul 0.359}          & 0.292          & {\ul 0.302}          & 0.281          \\ \hline

\textbf{ENq-Bic}             & 0.330          & 0.341          & 0.276          & 0.259          & 0.318         
 \\ 
\textbf{ARq-Bic}             & 0.297          & 0.332          & 0.297          & 0.260          & 0.311          
\\
\textbf{Biq-ENc}             & {\ul 0.383}    & 0.339          & \textbf{0.349} & 0.275          & {\ul 0.354}    \\
\textbf{Biq-ARc}             & 0.343          & 0.349          & 0.307          & 0.283          & 0.336         
 \\
\textbf{Biq-Bic}             & 0.341          & 0.315          & 0.265          & 0.251          & 0.307
 \\ \hline
\textbf{Bilingual-train-all} & 0.328          & 0.324          & 0.278          & 0.283          & 0.317          \\
\textbf{4lingual-train-all}  & \textbf{0.404} & \textbf{0.414} & 0.296          & \textbf{0.319} & \textbf{0.357}
\\ \hline
\end{tabular}
\end{adjustbox}
\caption{Multilingual evaluation MRR@10.}
\label{tab:table 2}
\end{table}

\begin{table}[t]
\begin{adjustbox}{width=\columnwidth,center}
\begin{tabular}{lcccc}
\hline
\textbf{Models}                       & \textbf{AR-UR}    & \textbf{UR-AR}    & \textbf{AR-EN}   & \textbf{EN-AR}  
\\ \hline
\textbf{XLMR-EN}       & 0.175          & 0.04                 & 0.097    & 0.019
\\ \hline
\textbf{EN-monolingual}             & 0.323          & 0.284                 & 0.247   & 0.284
\\
\textbf{AR-monolingual}    & 0.366	& 0.352		& 0.352	&0.36      \\ \hline
\textbf{ENq-ARc}                & 0.225	& 0.342		& 0.2	
& 0.342      \\
\textbf{ARq-ENc}          & \textbf{0.446}	& 0.34	    	& 0.357	
& 0.277      \\ \hline
\textbf{ENq-Bic}            & 0.34 	& 0.344	& 0.272	&0.369
 \\ 
\textbf{ARq-Bic}           & 0.376	& 0.368	     & \textbf{0.382}	 
& 0.337\\
\textbf{Biq-ENc}          & {\ul 0.424}	& {\ul 0.385}	& {\ul 0.37}	
& \textbf{0.423}
\\
\textbf{Biq-ARc}        &0.409	           & 0.368	          	& 0.341	& 0.343\\
\textbf{Biq-Bic}         & 0.369	           & \textbf{0.39}	          	& 0.36	
& {\ul 0.383}\\ \hline
\textbf{Bilingual-train-all}  & 0.349	& 0.329		            & 0.328	& 0.316       \\
\textbf{4lingual-train-all}       &0.207	& 0.099		               & 0.32 &0.281   
\\ \hline
\end{tabular}
\end{adjustbox}
\caption{Cross-lingual evaluation MRR@10.}
\label{tab:table 3}
\end{table}

\section{Training Approaches of Multilingual Retrieval Models}
For retrieval, we employ a dense retrieval approach \citep{karpukhin-etal-2020-dense} using the sentence transformer framework that adds a pooling layer on top of LLM embeddings and produces fixed-sized sentence embedding \citep{reimers-gurevych-2019-sentence}. 
The loss function is designed within the framework of contrastive learning, which helps create an embedding space that brings related queries and their relevant passages closer together while pushing away queries and irrelevant passages \citep{Oord-Contrastive}, and formally defined as:

\begin{align*}
J_{\mathrm{CL}}&(\theta) = \\-\frac{1}{M}\sum_{i=1}^M&\log\frac{\exp{\sigma(f_{\theta}(x^{(i)}), f_{\theta}(y^{(i)}))}}{\sum_{j=1}^M\exp{\sigma(f_{\theta}(x^{(i)}), f_{\theta}(y^{(j)}))}}
\end{align*}

where $\sigma$ is a similarity function (a cosine similarity), $f_{\theta}$ is the sentence encoder, $\{x^{(i)}, y^{(i)}\}_{i=1}^M$ (where \emph{M} is batch size) are positive labels and other in-batch examples treated as negative \citep{henderson2017efficient, gillick-etal-2019-learning, karpukhin-etal-2020-dense}.

We explore four different training approaches:

(1) \textbf{Monolingual training}, in this method, both the query and the passages are in the same language $L{_i}$.

(2) \textbf{Cross-lingual training} exploits a pair of languages during training in a traditional way; while queries are in the language $L{_i}$, passages are in the language $L{_j}$.

(3) \textbf{Mixed approach}. In this strategy, we construct the training data by combining monolingual and cross-lingual methods. There are three specific ways we develop the training samples:

- \textbf{Monolingual queries with bilingual collection}: Half of the passages in the collection are in the same language as the queries $L{_i}$, while the other half is in a different language $L{_j}$.

- \textbf{Bilingual queries with monolingual collection}: Here, the queries are presented in two different languages, $L{_i}$ and $L{_j}$, while the collection consists of passages in only one language, $L{_i}$.

- \textbf{Bilingual queries with bilingual collection}: In this case, both the queries and the passages can be in the languages $L{_i}$ and $L{_j}$.

One of the main differences between cross-lingual training and mixed training is that in cross-lingual training, the queries and passages are always in different languages. In contrast, mixed training allows for the query language to be either in the same language as the passage language ($L{_i}$) or a different language ($L{_j}$). We hypothesize that a mixed approach can enhance the diversity of training examples and improve cross-lingual interaction between languages.

(4) \textbf{Translate-train-all approach}: This approach involves training different translations of the training dataset simultaneously. In the previous mixed approach, queries or the collection are evenly divided between two languages. In this training mode, we expand the collection by adding another translation. 
Translate-train-all resembles monolingual training in structure (same-language pairs) but improves language coverage by training in multiple languages simultaneously.

\begin{figure*}[t]
\centering
  \includegraphics[width=\textwidth]{{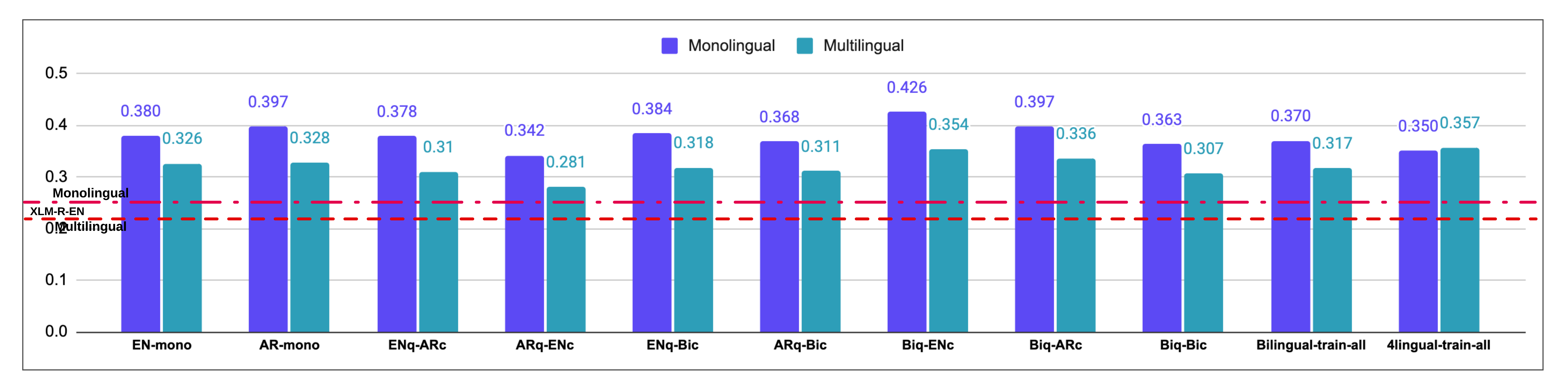}}
  \caption{Comparison of the model performance averaged across languages between multilingual and monolingual evaluations. Red lines are the performance of a baseline model XLM-R-EN.}
  \label{fig:figure 1}
\end{figure*}

\section{Experimental Setup}
The variations of the settings described above can create numerous combinations depending on how many languages are involved in the experiment.
For monolingual, cross-lingual, and mixed training approaches, we focus on using  English and Arabic. For two main reasons: English is the primary language of the XLM-R\textsubscript{Base} model, and the language of the MS MARCO dataset, and Arabic is the language that was mainly used for domain adaptation. For the translate-train-all approach, we experiment with four languages: English, Arabic, Urdu and Russian.

\subsection{Datasets}
For training, we use the MS MARCO \citep{Bajaj-MSMARCO}, a large-scale English-language dataset widely adopted for training and evaluating dense retrieval models. It contains over 500,000 real-world queries paired with a collection of 8.8 million passages; \citet{Bonifacio-mmarco} released machine-translated variants of MS MARCO for 13 languages, including Arabic and Russian (the collection was not translated into Urdu). Applying machine translation, we translated MS MARCO into Urdu to include this language in the experiment.
For evaluation, we combined the train and development splits of the QRCD (Qur'anic Reading Comprehension Dataset) \citep{malhas2020ayatec} to increase the size of the test set, resulting in 169 queries used for evaluation. The answers provided are exhaustive, meaning all Qur’anic verses directly responding to the questions have been thoroughly extracted and annotated. The queries reflect contemporary, real user information needs, so they are both valid and salient today. This pairing of complete passage coverage with today’s intents enables rigorous IR evaluation while preserving direct applicability to current user scenarios.
The language of the QRCD dataset is Arabic; to evaluate in other languages, we use verified translations of this dataset to English, Russian, and Urdu. 
We use the Holy Quran text (Arabic), Sahih International translation (English), Elmir Kuliev (Russian), and Ahmed Raza Khan (Urdu) as retrieval collections.\footnote{\url{https://tanzil.net/trans/}}

\begin{figure*}[t]
  \includegraphics[width=\textwidth]{{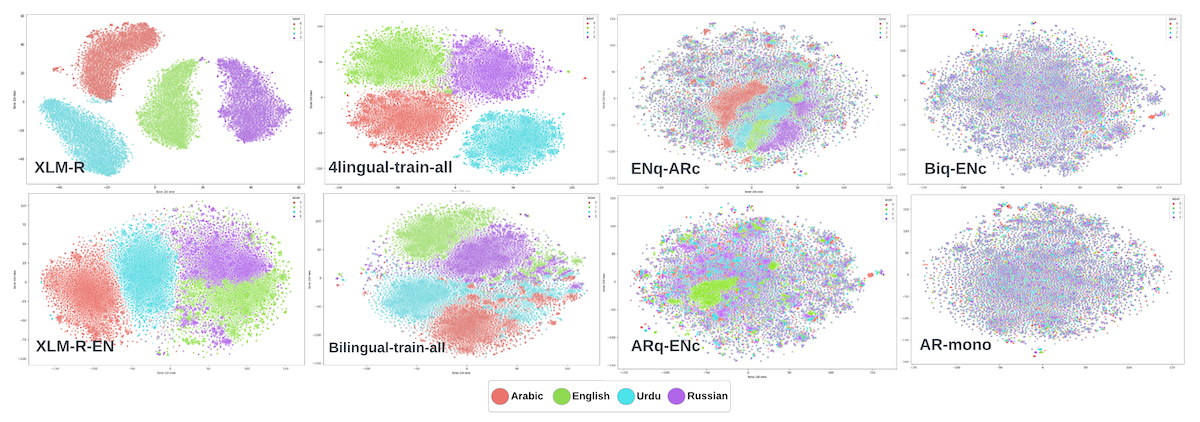}}
  \caption{2D t-SNE images of the representation of the Quranic verses embedding space in four languages.}
  \label{fig:figure 2}
\end{figure*}

\subsection{Evaluation Approach, Metrics and Baseline}
We evaluate our methods in three different settings: monolingual, cross-lingual, and multilingual.
In a monolingual setting, we evaluate using four languages.
For the cross-lingual evaluation, we analyze two language pairs: Arabic and Urdu, which share similarities in writing systems and vocabulary, and Arabic and English, which represent linguistically distinct languages.
In the multilingual evaluation, we combine collections of Quranic texts in four different languages. We then assess retrieval performance for each language by varying the query languages.

We use the MRR@10 (Mean Reciprocal Rate), the official evaluation metric of the MS MARCO dataset, as the main metric.

As a baseline, we train the XLM-R\textsubscript{Base}  in a monolingual setting using English MS MARCO (XLM-R-EN model) that allows us to see the effects of domain adaptation. For the rest of the retrieval models described below, we use the XLMR-4-ID model for training.

\begin{figure*}[t]
\centering
  \includegraphics[width=\textwidth]{{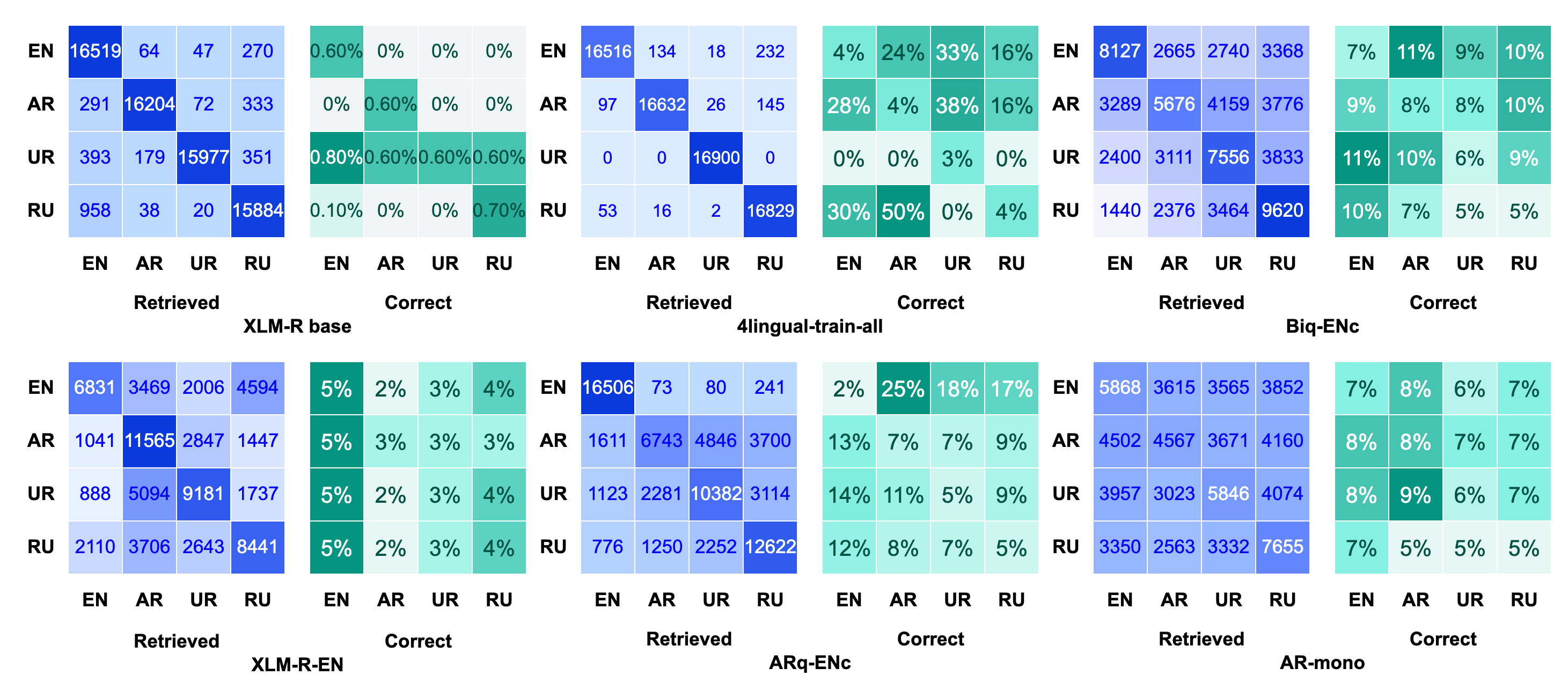}}
  \caption{The heat map in blue shows how many passages from what language collection were retrieved. The heat map in green shows the percentage of correct passages out of retrieved passages. Language collection is the x-axis, and queries in a specific language are the y-axis.}
  \label{fig:figure 3}
\end{figure*}

\begin{table*}[]
\begin{adjustbox}{width=\textwidth,center}
\begin{tabular}{c|c|ccc}
\hline
\textbf{Model(s)} & \textbf{Three Large Models} & \multicolumn{3}{c}{\textbf{Single Versatile Lightweight Model}}                                      \\ \hline
\textbf{Compute Substrate} & \textbf{24GB A10 GPU}     & \textbf{12 GB RTX 3080TI} & \textbf{CPU EM-A210R-HDD 16GB} & \textbf{Lambda Function} \\
\textbf{Provider}          & Lambda Labs               & Vast.ai                   & Scaleway                       & AWS                      \\
\textbf{VRAM / RAM}        & 24 GB                     & 12 GB                     & 16 GB                          & 2 GB+                    \\
\textbf{MRC}               & $\sim$ USD 540           & $\sim$ USD 150           & $\sim$ USD 48                 & $\sim$ USD 10 - USD 20     \\ \hline
\end{tabular}
\end{adjustbox}
\caption{MRC (monthly recurring cost) comparison from multiple compute substrate providers.}
\label{tab:Table 4}
\end{table*}

\begin{table*}[]
\begin{adjustbox}{width=\textwidth,center}
\begin{tabular}{ccccc}
\hline
\textbf{}               & \textbf{Metric}    & \textbf{MENA/EU Region} & \textbf{North America Region} & \textbf{APAC} \\ \hline
\multirow{3}{*}{Before} & Avg. Latency       & 40.93ms                 & 204.53ms                      & 772.01ms      \\
                        & Median P50 Latency & 38.56ms                 & 206.03ms                      & 765.09ms      \\
                        & P95 Latency        & 89.01ms                 & 397.87ms                      & 1380.37ms     \\ \hline
\multirow{3}{*}{After}  & Avg. Latency       & 25.33ms                 & 160.03ms                      & 408.99ms      \\
                        & Median P50 Latency & 23.69ms                 & 150.73ms                      & 402.66ms      \\
                        & P95 Latency        & 48.99ms                 & 220.27ms                      & 980.11ms      \\ \hline
\%DELTA                 & P50 (lower better) & -38.6\%                 & -26.8\%                       & -47.4\%       \\ \hline
\end{tabular}
\end{adjustbox}
\caption{RUM latency (ms) by region before/after model deployment; lower is better.}
\label{tab:Latency}
\end{table*}

\subsection{Results}
In all tables, the best score is in bold, and the second-best is underlined. In the monolingual evaluation (Table ~\ref{tab:table 1}), the model with the highest average performance across languages (0.426) is the  Biq-ENc (Bilingual Queries English Collection), which was trained with a mixed approach. It also demonstrates the best performance in both Arabic and Urdu. 
In the multilingual evaluation (Table ~\ref{tab:table 2}), the best-performing model on average (0.357) is the 4lingual-train-all (translate-train-all approach). The Biq-ENc model achieves the highest score (0.349) in Urdu  and also second best for average performance across four languages (0.354).
As illustrated in Figure ~\ref{fig:figure 1}, all models outperform the baseline on average across languages. 
In the cross-lingual evaluation (Table ~\ref{tab:table 3}), the Biq-ENc model is the top performer (0.423) for the EN-AR pair; for all other pairs, this model has the second-best results.
Overall, the results indicate that the mixed training approach yields promising outcomes, with the Biq-ENc model consistently demonstrating strong performance across all evaluation types: monolingual, multilingual, and cross-lingual.

\section{Analysis}
To assess how training strategies shape the multilingual embedding space, we apply the t-SNE algorithm \citep{Maaten2008VisualizingDU} (Figure~\ref{fig:figure 2}). To evaluate each model's cross-lingual capability, we examine heat maps for retrieval (blue) and correctness (green), which highlight monolingual (diagonal) and cross-lingual (off-diagonal) retrieval ability (Figure~\ref{fig:figure 3}).
To juxtapose models that demonstrate strong cross-lingual ability with those that remain monolingually biased, we include the XLM-R\textsubscript{Base} model, which is not fine-tuned for retrieval.
The first t-SNE image in the upper row (Figure~\ref{fig:figure 2}) shows that XLM-R\textsubscript{Base} produces four distinct language clusters and retrieves passages almost exclusively from the same language as the query. However, its accuracy remains extremely low, suggesting minimal ability for cross-lingual retrieval.
The 4lingual-train-all (translate-train-all approach) model also clusters by language and is biased toward retrieving from the same language. However, it achieves a higher percentage of correct answers, including some off-diagonal results.
The XLM-R-EN model (trained on English MS MARCO from XLM-R model without domain adaptation) produces less pronounced clusters, with partial overlap between English and Russian as shown in the t-SNE image. Remarkably, this model performs well on English and Russian datasets, but poorly on Urdu and Arabic. The green heat map aligns with language collections (vertical alignment), with the darkest column corresponding to English passages, which is expected given the model's English-centric training.
The ARq-ENc model (Arabic queries, English collection) shows a more unified embedding space, though English cluster remains somewhat distinct. The blue heat map shows the darkest quadrant in the English-English retrieval, yet with a very low percentage of correct answers. Notably, this model performs poorly, particularly on English. 

Conversely, t-SNE plots of Biq-ENc (bilingual queries, English collection) and AR-mono (domain-adapted XLM-R trained on Arabic MS MARCO) show a more homogeneous structure without clear language clusters. The heat maps are more uniformly colored for both retrieved and correct answers. Their off-diagonal results highlight stronger cross-lingual ability, supporting the idea that domain adaptation and mixed training encourage models to learn embedding space akin to language-agnostic representations, leading to improved performance in multilingual retrieval.

\section{Deployment Considerations}
 Table ~\ref{tab:Table 4} demonstrates possible deployment considerations and compares costs for deploying three separate retrieval models, each around 1GB (trained from XLM-R\textsubscript{Base}), versus deploying one lightweight model of 400 MB in size (e.g. Biq-ENc model). The table shows that deploying one smaller model allows a reduction of costs by about 70\% when deploying on a GPU-based server. Lower memory consumption and faster loading of a single lightweight model allow us to consider the deployment option on CPU-based servers, which further cut the cost by 70\%. Python's runtime overhead, garbage collection, and large dependencies introduce inefficiencies in memory utilization, increasing overall deployment size. Leveraging Rust language capabilities to eliminate Python's inefficient memory management enables to reduce overall memory consumption to 30-50\%, which can pave the way for deployment on compact serverless runtimes such as AWS Lambda functions, presenting the most cost-effective and scalable solution, potentially reducing monthly recurring costs to as low as USD 10 - USD 20.

\section{Production Performance}
We evaluate end-to-end latency with real-user monitoring (RUM) before and after deploying the new model (see Table~\ref{tab:Latency}). Measurements reflect browser-observed round-trip times from production traffic, exclude known bots, and are reported by region at three cut points: mean, median (P50), and tail (P95). As shown in Table~\ref{tab:Latency}, latency dropped across all regions after deployment. Median latency decreased by 38.6\% in MENA/EU (38.6$\to$23.7 ms), 26.8\% in North America (206.0$\to$150.7 ms), and 47.4\% in APAC (765.1$\to$402.7 ms). Means showed similar gains ($-38.1\%$, $-21.8\%$, $-47.0\%$). Tail latency (P95) improved sharply: MENA/EU $-45.0\%$ (89.0$\to$49.0 ms), North America $-44.7\%$ (397.9$\to$220.3 ms), and APAC $-29.0\%$ (1380.4$\to$980.1 ms). Lower tail latency materially improves perceived responsiveness under load and on slower networks. The APAC tail remains higher due to network distance; further reductions will require geographic routing and edge capacity in addition to model-side efficiency. The Appendix~\ref{appendix:b} provides additional details on user activity and other extrinsic evaluations, including the assessment of real-user queries.

\section{Related work}
Recent work in cross-lingual and multilingual information retrieval (CLIR and MLIR) explores extending monolingual dense retrievers such as ColBERT to the multilingual setting, often using XLM-R\textsubscript{Base} as the backbone encoder \citep{Nair2022TransferLA, Lawrie2022NeuralAT}. Our approach differs in that we use XLM-R\textsubscript{Base} within a sentence embedding framework, which is more latency-efficient and scalable for real-world deployment. Unlike ColBERT-style late interaction models, we adopt full sentence representations with in-batch negatives—an approach shown to yield strong performance with lower computational cost \citep{qu-etal-2021-rocketqa, ren-etal-2021-rocketqav2, karpukhin-etal-2020-dense}.
Multilingual sentence embeddings have also been actively studied, with methods like LaBSE \citep{feng-etal-2022-language}, mSimCSE \citep{hu-etal-2023-language}, LASER \citep{artetxe-schwenk-2019-margin}, and multilingual variants of SBERT \citep{reimers-gurevych-2020-making} demonstrating strong performance across a variety of tasks. While many of these focus on general-purpose representation learning, our work specifically investigates how different training configurations affect multilingual retrieval quality and embedding space alignment in a domain-specific setting.

\section{Conclusion}
Our proposed mixed training approach has shown promising results across all evaluation settings, highlighting its beneficial properties for use in MLIR systems that need to handle various retrieval scenarios. Furthermore, the efficiency gained by deploying a single lightweight and versatile model proves to be a superior option for balancing performance, affordability, and scalability.

\section*{Limitations}
Despite including different types of languages in our experiment and adding low-resource ones like Urdu, the results may vary with a significantly larger number of languages. Additionally, our mixed and cross-lingual training setups rely on parallel corpora, which may not generalize well to settings where such resources are unavailable or noisy. Finally, although we deployed a lightweight model, performance and efficiency trade-offs on truly resource-constrained devices (e.g., mobile or edge environments) remain to be fully explored.

\section*{Ethical Considerations}
This work involves the retrieval of religious texts, specifically the Holy Quran, which holds deep cultural and spiritual significance for millions of people. We have taken care to use verified and official translations of Quranic text. However, variations in translation style and theological interpretation may still impact how passages are retrieved and understood across languages.
We acknowledge the responsibility that comes with building search tools in sensitive domains. The system is designed to assist in information access, not to provide religious or legal rulings. Care should be taken in downstream use cases, particularly in educational, interfaith, or legal contexts. Additionally, as with any multilingual system, there remains a risk of uneven performance across languages, which could inadvertently prioritize or marginalize certain linguistic groups.
We recommend future work consider input from domain experts, theologians, and community stakeholders to guide responsible deployment, especially when extending the system to broader religious or cultural corpora.

\section*{Acknowledgment}
We thank the Program Chairs of the EMNLP Industry Track for their continuous support of research on underrepresented domains. We also gratefully acknowledge the Dubai AI Campus for providing the computing infrastructure that made this work possible. We extend our appreciation to our colleagues and collaborators for their support and inspiration throughout this project.

\bibliography{anthology,custom}

\clearpage

\appendix
\section{Appendix}
\label{sec:appendix}

Our reduction method consists of the following steps:
\begin{enumerate}
    \item We select English, Arabic, Russian, and Urdu texts from a multilingual variant of the C4 corpus \citep{Raffel2019ExploringTL} and train and SentencePiece BPE tokenizer.
    \item  Find the intersection between the new tokenizer and the XLM-R\textsubscript{Base} tokenizer\footnote{\url{https://huggingface.co/FacebookAI/xlm-roberta-base}}, the tokens inside of intersection and the corresponding weights will be selected for the new embedding matrix of the XLM-R4 model (34k tokens).
    \item The encoder weights from XLM-R\textsubscript{Base} get copied to the new XLM-R4 model as is.

\end{enumerate}
Evaluation of XLMR-4 on the XNLI dataset demonstrates only a slight drop in performance (around 1.77\% across all languages) compared to the XLM-R\textsubscript{Base} (see Table ~\ref{tab:XNLI comparison}). At the same time, we manage to significantly reduce the number of parameters by trimming the embedding matrix (EM) (see Table ~\ref{tab:model weights}).

The domain adaptation of XLM-R4 takes the following steps:
\begin{enumerate}
    \item We train a new SentencePiece BPE tokenizer using a multilingual Islamic Corpus and find the intersection between the new Islamic tokenizer and the XLM-R4 tokenizer. All the tokens outside of the intersection (9k tokens) are added to the embedding matrix of the XLMR-4 model, and the weights for new tokens are assigned by averaging existing weights of subtokens from the XLM-R4 model. 
    \item We continue pre-training XLM-R4 using the domain-specific corpus which gives us the XLM-R4-ID (Islamic domain) model. For more details on the hyperparameters, refer to Appendix~\ref{sec:appendix}.
\end{enumerate}

\begin{figure*}[!htbp]
\centering
  \includegraphics[width=\textwidth]{{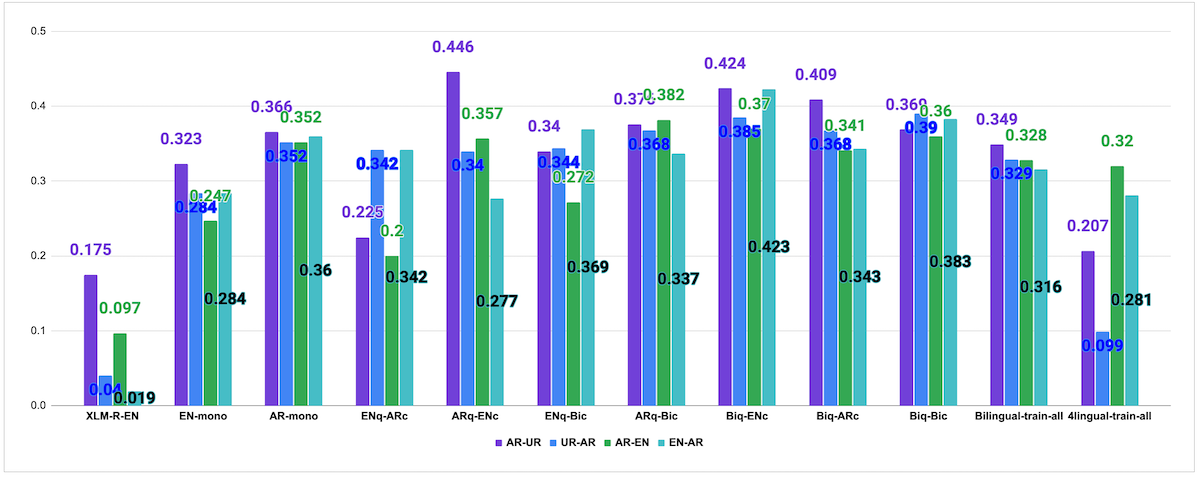}}
  \caption{Comparison of model performance on cross-lingual evaluation.}
  \label{fig:figure 4}
\end{figure*}

\begin{table}[h]
\begin{adjustbox}{width=\columnwidth,center}
\begin{tabular}{cccc}
\hline
Model     & Size   & \#params & EM    \\ \hline
mBERT     & 714 MB & 178 M    & 92 M  \\
XLM-R\textsubscript{Base} & 1.1 GB & 278 M    & 192 M \\
XLM-R4    & 481 MB & 119 M    & 33M   \\ \hline
\end{tabular}
\end{adjustbox}
\caption{Comparison of models' size}
\label{tab:model weights}
\end{table}

\begin{table}[h]
\begin{adjustbox}{width=\columnwidth,center}
\begin{tabular}{ccccc}
\hline
\textbf{Model}        & \textbf{en}             & \textbf{ru}             & \textbf{ar}             & \textbf{ur}             \\ \hline
XLM-R\textsubscript{Base}    & \textbf{84.19} & \textbf{75.59} & \textbf{71.66} & \textbf{65.27} \\
XLM-R4       & {\ul 83.21}    & {\ul 72.75}    & {\ul 70.48}    & {\ul 64.95}    \\
mBERT        & 82.1           & 68.4           & 64.5           & 57             \\
mBERT 15lang & 82.2           & 68.7           & 64.9           & 57.1           \\
DistillmBERT & 78.5           & 63.9           & 58.6           & 53.3           \\ \hline
\end{tabular}
\end{adjustbox}
\caption{Results on cross-lingual transfer for four languages of the XNLI dataset. XLM-R\textsubscript{Base} and XLM-R4 results are averaged over five different seeds.}
\label{tab:XNLI comparison}
\end{table}

\begin{table}[h]
\begin{adjustbox}{width=\columnwidth,center}
\begin{tabular}{ccc}
\hline
\textbf{Computing Infrastructure}    & \multicolumn{2}{c}{1x H100 (80 GB)}    \\ \hline
\multicolumn{2}{c}{\textbf{Hyperparameter}} & \textbf{Assignment}                     \\ \hline
\multicolumn{2}{c}{number of epochs}        & 60 \\ \hline
\multicolumn{2}{c}{batch size}              & 128                             \\ \hline
\multicolumn{2}{c}{maximum learning rate}   & 0.0005                        \\ \hline
\multicolumn{2}{c}{learning rate optimizer} & Adam                                    \\ \hline
\multicolumn{2}{c}{learning rate scheduler} & None or Warmup linear                   \\ \hline
\multicolumn{2}{c}{Weight decay}            & 0.01                                    \\ \hline
\multicolumn{2}{c}{Warmup proportion}       & 0.06                                    \\ \hline
\multicolumn{2}{c}{learning rate decay}     & linear                                  \\ \hline
\end{tabular}
\end{adjustbox}
\caption{Hyperparameters for pre-training of XLM-R4-ID model.}
\label{tab:appendix-table-a}
\end{table}

\begin{table}[!htbp]
\begin{adjustbox}{width=\columnwidth,center}
\begin{tabular}{ccc}
\hline
\textbf{Computing Infrastructure} & \multicolumn{2}{c}{1x H100 (80 GB)} \\ \hline

\multicolumn{2}{c}{\textbf{Hyperparameter}}     & \textbf{Assignment}           \\ \hline
\multicolumn{2}{c}{number of epochs}            & 10                            \\ \hline
\multicolumn{2}{c}{batch size}                  & 256                             \\ \hline
\multicolumn{2}{c}{learning rate}               & 2e-5                          \\ \hline
\multicolumn{2}{c}{pooling}                     & mean                         \\ \hline

\end{tabular}
\end{adjustbox}
\caption{Hyperparameters for training retrieval models.}
\label{tab:appendix-table-c}
\end{table}

\section{Real User Metrics}\label{appendix:b}
During August 2025, the deployed system processed 84,255 requests originating from 2,630 unique IP addresses (median $\approx$ 32 requests/IP), with a total data volume of approximately 1.9 GB served. The cache hit rate by bytes was approximately 11\%, consistent with a predominantly dynamic, search-intensive workload. Request counts exclude known automated traffic through Cloudflare bot-score and hosting-ASN filters. Clear diurnal usage patterns, distribution across residential and mobile ASNs, and a balanced mix of HTML, asset, and API requests all indicate genuine end-user activity. The system operated in CPU-only serving mode with an uptime of $\geq$99.97\%. (All metrics are aggregated and privacy-preserving.)

On the user-facing site over the same period (via real-user monitoring, RUM), Success@5 was approximately 58–62\% ($\tau$ = 15 s; results page viewed $\geq$15 s with top-5 results rendered). Abandonment was approximately 18–22\%, defined as either no top-5 impression, dwell time <15 s, or a query reformulation within 30 s.

\textbf{Human evaluation.} We constructed a parallel set of 25 real-user queries in AR/UR/RU/EN. For each language, two independent annotators rated all retrieved results on a 3-point relevance scale (0 = irrelevant, 1 = partially relevant, 2 = highly relevant). Inter-annotator agreement was substantial (weighted Cohen’s $\kappa$ $\approx$ 0.61). Compared to an XLM-R Base baseline, our model achieved consistent gains in nDCG@10 of +0.06 - +0.10 across all four languages. Example queries are provided in Table~\ref{tab:User-queries}.

\begin{table}[h]
\centering
\begin{tabular}{c p{0.9\linewidth}}
\hline
\textbf{Query ID} & \textbf{Query} \\ \hline
1  & How is the universe created? \\
2  & What is the purpose of life of man on earth? \\
3  & How is the fetus formed in the womb? \\
4  & What function does the frontal lobe of the brain have? \\
5  & How is the rain created? \\
6  & What is the condition at the depth of the sea? \\
7  & Why do mountains stand still on the surface of the earth? \\
8  & Can animals communicate in their own languages? \\
9  & Will we be held accountable for our deeds? \\
10 & What is Hijab? \\
11 & Will the world come to an end and how will it happen? \\
12 & Who was Jesus? \\
\hline
\end{tabular}
\caption{Examples of real user queries}
\label{tab:User-queries}
\end{table}

\end{document}